# Radio Core-Dominance of *Fermi*-Blazars: Implication for Blazar Unification

**E.U. Iyida***, **F.C. Odo, A.E. Chukwude**

Astronomy and Astrophysics Research Group, Department of Physics and Astronomy, University of Nigeria, Nsukka, Department of Physics and Astronomy, Faculty of Physical sciences, University of Nigeria, Nsukka, Nigeria

*email: uzochukwu.iyida@unn.edu.ng

**Abstract**

In this paper, we use the distributions of two-component radio emissions to investigate the consequences of relativistic beaming on the unified scheme of blazar populations and radio galaxies. Our results show remarkable continuity in the distributions of source luminosities (radio core and extended luminosities, $L_C$ and $L_E$ respectively as well as γ-ray luminosity $L_\gamma$) from radio galaxies at low luminosities to FSRQs at high luminosities through BL subclasses as expected in the blazar unification scheme in a sense suggesting that the sequence of BL Lacs, FSRQs and radio galaxies represents progressively misaligned populations of AGNs. Distribution of radio core-dominance is consistent with average projection angle of 13.5°, 14.8°, 16.8°, 20.4° and 28.2° for ISPs, LSPs, FSRQs, HSPs, and radio galaxies, respectively. Linear regression analyses of our data yield significant anti-correlation ($r > 0.60$) between core-dominance parameter ($C$) and $L_E$ in each individual subsample: the correlation is significant only when individual subsamples are considered. There is a systematic sequence of the distribution of the different subclasses on the $C - L_E$ plane. Nevertheless, little or no correlation between $C$ and $L_C$ or between $C$ and $L_\gamma$ ($r < 0.50$) was observed. There is a clear dichotomy between high synchrotron-peaking BL Lacs and other BL Lac subclasses. The results are consistent with a unified view for blazars and can be understood in terms of relativistic beaming persisting at largest scales.

## 1. Introduction

The central regions of active galaxies that shine much better than the combined light from the entire galaxy are called Active Galactic Nuclei (AGNs). These AGNs have very high luminosity (between $10^{42}$ - $10^{48}$ erg/s) in a small volume of space (probably >>1pc$^3$) via accretion processes within the supermassive black hole (SMBH) (Lobanov et al. 2006; Angelakis et al. 2015). AGNs play essential roles in the formation and evolution of galaxies. Of all the AGNs, about 15 % are radio-loud with radio-loudness factor $R_L$ defined as

$$R_L = \log\left(\frac{S_{5GHz}}{S_B}\right) \geq 10 \tag{1}$$

while the remaining 85 % are radio-quiet AGNs (Xu et al., 1999; Balokovic et al., 2012). Blazars are particular subclass of radio-loud AGNs whose broadband radiations are mainly controlled by non-thermal emissions that is produced by the relativistic plasma jet which is allied straight to the observer. Based on spectroscopic properties, blazars are categorized into BL Lacertae Objects (BL Lacs) and flat spectrum radio quasars (FSRQs). However, the luminosity of the broad-line region (BLR) is distinctive between blazars (see e.g., Ghisellini and Celotti, 2001). Ghisellini et al., (2011) used the ratio of broad emission line luminosity to Eddington luminosity ($L_{BLR}/L_{Edd}$) to classify blazars. They pointed out that FSRQs have $L_{BLR}/L_{Edd} \geq 5 \times 10^{-4}$, while BL Lacs have $L_{BLR}/L_{Edd} < 5 \times 10^{-4}$. This division between FSRQs and BL Lacs may imply that they have different accretion regime (Sbarrato et al. 2014). Furthermore, BL Lacs are classified based on the synchrotron peak frequency $v_{s,peak}$ of their broadband spectral energy distributions (SEDs). Thus, there are high synchrotron (HSPs), intermediate synchrotron (ISPs) and low synchrotron (LSPs) peaked BL Lacs with synchrotron peak frequencies in the range $\log v_{peak}^{syn} > 15 (\text{Hz})$, $14 \leq \log v_{peak}^{syn} \leq 15 (\text{Hz})$ and $\log v_{peak}^{syn} < 14 (\text{Hz})$ respectively (Ackermann et al., 2015; Fan et al. 2016; Ajello, 2020). Previous studies have shown that blazars are very strong emitters of γ-rays (GeV up to TeV) and the radiations from these objects are characterized by apparent superluminal motion in their parsec-scale relativistic jets, large and rapid variability as well as strong polarization (Andruchow et al. 2005; Padovani, 2007; Abdo et al. 2010; Urry, 2011; D'Abrusco et al., 2019; Abdollahi et al., 2020). All these observational properties are attributed to relativistic beaming effect at small viewing angles (Urry and Padovani, 1995).

In a two-component jet model, the total luminosity ($C_T$) from an AGN is the sum of the core ($L_C$) and extended ($L_E$) components and the ratio ($C$) of $L_C$ to $L_E$, known as the core-dominance parameter is often used as a measure of relativistic beaming (Orr and Browne, 1982; Fan and Zhang, 2003). In the core-dominated sources, $C$ is large (>1) due to Doppler boosting of the core emission, while in lobe dominated part, $C$ is less pronounced (<1) since the extended emission is supposedly isotropic (Urry and Padovani, 1995). Thus, many properties of blazars are associated with relativistic beaming effects once seen at a very small angle to the line of sight. It is as a result of this exceptional viewing angle in addition to the depletion of collimated relativistic elements within the line of sight that results in the observation of blazars (Dennert et al. 2000). Due to increased observations and better theoretical understanding of the nature of these sources, a unification scheme has been proposed that fundamentally attempts to explain the relationship between the blazar subclasses through blazar sequence (e.g. Fossati et al. 1998). Based on broadband spectral energy distribution, the blazar sequence is a scenario in which the bolometric luminosity of blazar governs the appearance of its spectral energy distribution. The most prominent result is the significant negative correlation between the synchrotron peak frequencies $v_{s,peak}$ and the synchrotron peak luminosities ($L_{syn}$) of the blazar population. The study of this relationship has been an unresolved problem in blazar investigation in a couple of decades and is still under discussion (see, e.g., Fossati et al., 1998; Ghisellini et al., 1998; Ghisellini and Tavecchio, 2008; Meyer et al. 2011; Ackermann et al., 2015; Mao et al. 2016; Nalewajko and Gupta, 2017; Iyida et al., 2019; Palladino et al., 2019; Pei et al., 2020c). It has been argued that the inclination of the relativistic jets to the line of sight pointedly gives rise to differences in the properties of diverse classes of blazars due to variations in relativistic beaming effect at varying angles (see, e.g., Vagnetti et al., 1991; Finke, 2013). Nonetheless, the assumption that BL Lac objects represent FSRQs with highly boosted continuum appears somewhat questionable since it was discovered that the amount of intrinsic power and relativistic beaming are higher in majority of FSRQs than in BL Lac objects (e.g. Padovani 1992; Chen et al., 2016; Ghisellini et al., 2017; Odo et al. 2017), thus, indicating that there are inherent differences between them.

However, since the community consensus now appears to favour a unification scheme in which FSRQs and BL Lacs are believed to be beamed counterparts of high and low luminosity radio galaxies, respectively (see, eg., Urry and Padovani, 1995; Ghisellini et al. 2009; Sbarrato et al. 2014), orientation connections between blazar subclasses as well as between blazars and radio galaxies are regaining attention with interesting results (Fan et al. 2011; Pei et al. 2019, Iyida et al. 2020). In particular, recent investigations in the *Fermi* era suggest that blazar sequence can be extended to radio galaxy populations by invoking an orientation sequence (e.g. Iyida et al., 2020). Also, Pei et al., (2020c) found that the orientation-dependent core-dominance parameter is strongly anti-correlated with luminosity distribution in a sample of quasars and Seyfert galaxies and argued for a connection between the core-dominance parameter and the spectral indices, which is often used as a parameter for investigating blazar sequence. Nevertheless, the discovery of some FSRQs with very low energies (see, e.g., Perlman et al., 1998; Raiteri and Capetti, 2016) and BL Lacs with high energies in γ-ray band detected by the *Fermi*-LAT (e.g. Ackermann et al., 2015) appears to break the simple dichotomy between traditional low and high luminosity sources and suggests that BL Lacs and FSRQs are a continuum in distributions of observational properties.

Although relativistic beaming of core emission provides a natural explanation for the extreme properties of blazars, compared to their supposedly parent populations of radio galaxies, this scenario appears to be substantially complicated by the apparent beaming of extended radio emission detected among traditional samples of BL Lacs and FSRQs (Qin and Xie, 1998; Odo and Ubachukwu, 2013; Odo et al., 2015). However, preliminary investigation of γ-ray emission from 80 *Fermi* detected blazars (Liu et al., 2016) yields a strong correlation between radio core-dominance and $\gamma$-ray luminosity, suggesting that relativistic beaming and orientation effects play a significant role in detection of $\gamma$-ray emissions of AGNs (see, also, Chen et al. 2016; Pei et al. 2020b). Thus, both relativistic beaming and blazar sequence hypotheses in these sources need to be closely re-examined in large and up-to-date samples. Partly motivated by these recent findings, in this paper, we use the radio core-dominance parameter to statistically investigate simple consequences of relativistic beaming hypothesis in the, so far, up-to-date sample of *Fermi*-LAT blazars and show statistically that relativistic beaming persists at largest scales in blazars.

## 2. Basic Theoretical Concepts

The source orientation and relativistic beaming theory postulates that extragalactic sources that are observed straight to the line of sight at a small angle are associated with relativistically boosted radio emissions from the core in relation to the lobes (Orr and Browne, 1982). This popular paradigm has remarkably explained the differences between various subsets of active galactic nuclei as related objects, the differences depending highly on the orientation and symmetry of the source as well as the observer's angle of sight. It has been reported (Jeyakumar et al., 2005; Chen et al., 2016, Odo et al., 2017; Resconi et al., 2017; Alhassan et al., 2019) that orientation as well as relativistic beaming is indispensable in explaining various emissions from blazar subclasses. A key statistical indicator of source orientation and relativistic beaming theory is the core-dominance parameter defined following Orr and Browne (1982) as

$$C = \left(\frac{L_C}{L_E}\right) = \frac{R_T}{2}\left[(1-\beta\cos)^{-n+\alpha} + (1+\beta\cos)^{-n+\alpha}\right] \quad (2)$$

where $R_T$ (0.024) is the ratio of core-extended flux densities when the core is viewed at $\varphi = 90°$, $n$ is a parameter that depends on the flow model: $n = 2$ for a continuous jet model, otherwise it is 3 if the emitting medium contains mainly blobs (Scheuer et al., 1979; Lind et al., 1985). However, for a continuous jet model with flat core of zero spectral index, equation (2) simplifies to

$$C = \left(\frac{L_C}{L_E}\right) = \frac{R_T}{2}\left[(1-\beta\cos)^{-2} + (1+\beta\cos)^{-2}\right] \quad (3)$$

Thus, it can be deduced from equation (2) that if $\beta \approx 1$ (see, e.g. Fan and Zhang, 2003) for relativistic jets, the average viewing angle can be approximated to

$$\varphi_m = \cos^{-1}\left[\frac{2C_m + R_T - \sqrt{R_T(8C_m + R_T)^{\frac{1}{2}}}}{2C_m}\right]^{\frac{1}{2}} \quad (4)$$

where $C_m$ is the source's median value. The equation (4) implies that the average viewing angle, $\varphi_m$ can be estimated from the distribution of $C$ if the $R_T$ is known.

Conversely, in the framework of the relativistic beaming effect at a small angle, the $C_T$ is related to the core-dominance parameter following Fan and Zhang (2003) as

$$\frac{C}{1+C} = \frac{L_C}{C_T} \quad (5)$$

Therefore, equation (5) implies that if $C \ll 1$ and $C_T$ is isotropic as in the case of relativistic beaming model, a positive relationship in the $C$ - $C_T$ plot is predicted. However, the $C$ - $L_C$ connection has a power law (see, e.g., Odo et al. 2015) of the form

$$\log C = \log C_b + \lambda \log L_c \quad (6)$$

$C_b$ is the smallest value of $C$ of the beamed sources, while $\lambda$ is its slope. This is real for a beamed source with boosted emission as a result of the relativistic Doppler effects due to the small viewing angle. However, if the extended emission is also affected for $\beta \approx 1$ and at a very small angle, then $(1-\beta\cos\varphi)^{-(n+\alpha)}$ becomes a constant. Thus, $L_E$ is simply expressed through equation (2) as

$$L_E = L_C(1+\cos\varphi)^{-(n+\alpha)} \qquad (7)$$

Therefore, this indicates that when the source is being viewed end-on, $L_E$ is beamed relative to $L_C$ as expressed in equation (2). The statistical implication is that the $L_E - L_C$ relation is strong as is expected in beamed sources whose angles of inclination are close to the line of sight. In effect, this defines $C$ as a quantity of interest in relativistic beaming with the source being increasingly detected end-on. Furthermore, several authors (e.g., Scheuer et al., 1979; Bridle et al., 1994; Fan and Zhang, 2003) and lately Pei et al. (2019; 2020a) have shown that $C$ is a very dependable statistical parameter for testing beaming hypothesis. In particular, it has been argued that $C$ is strongly related to γ-ray luminosity ($L_\gamma$) in samples of γ-ray-loud blazars in a general form (e.g. Wu et al. 2014; Pei et al. 2020b):

$$\log\left(\frac{L_\gamma}{L_E}\right) = \lambda \log(1+C), \qquad (8)$$

where λ is a sample dependent factor. In effect, the two-component jet model of radio – to γ-ray emission from AGNs leverages a correlation between radio core-dominance parameter and γ-ray luminosity and a strong correlation between the two parameters (e.g. Liu et al. 2016; Pei et al. 2020b) is an indication that γ-ray emission from blazars is strongly beamed. Thus, $C - L_\gamma$ correlations can also be used to study both relativistic beaming and blazar sequence hypotheses in samples of γ-ray loud sources (e.g. Pei et al. 2020b).

However, for a source with spectral flux density ($S_v$) at two observing frequencies $v_1$, $v_2$ and redshift ($z$), $C$ can be expressed following Fan and Zhang (2003) as

$$C = \left(\frac{L_C}{L_E}\right) = \frac{L^C_{v_2}(GHz)}{C^T_{v_1}(GHz) - L^C_{v_2}(GHz)} \left(\frac{v_1}{v_2}\right)^{-\alpha E} (1+z)^{-\alpha E} \qquad (9)$$

with $\alpha E$ being the energy spectral index from the extended component while the radio core spectral is assumed to be zero. A simple consequence of this is that if the wide range in distributions of $C$ and the two luminosity components ($L_C$ and $L_E$) is largely a beaming effect, a correlation is predicted between $C$ and $L$ for beamed radio sources in absence of redshift effect.

However, the *k-correction* term $(1+z)^{(\alpha-1)}$ in equation (9) implies that the luminosity properties of radio sources are induced by redshift effects such that $C - z$ dependence is envisaged. This is particularly common at high redshifts for these radio sources.

## 3. Data Analysis and Results

Analysis in this paper is based on the fourth catalogue of blazars detected by *Fermi* Large Area Telescope (*Fermi*-LAT) which contains 5098 objects in the energy range from 50 MeV to 1 TeV (Abdollahi et al. 2020; Ajello et al. 2020). From the catalogue, Pei et al. (2020b) compiled the radio core-dominance parameter of 4388 AGNs which includes 584 optically identified blazars, comprising 252 BL Lacs and 283 FSRQs. For BL Lacs in the sample, the γ-ray luminosity and observed radio information on core and extended flux-densities was obtained and we calculate the radio luminosity using $L = 4\pi d_L^2 \cdot S_{obs} (1+z)^{(\alpha-1)}$ with $\alpha$ being the energy spectral index defined as $S_\nu \approx \nu^{-\alpha_\nu}$ and taken as 1.0 and 0.0 for the extended and core emissions, respectively, while $d_L$ is the luminosity distance given (Alhassan et al., 2019) as $d_L = H_o^{-1} \int \left[ (1+z)^2 (1+\Omega_m z) - z(2+z)\Omega_\Lambda \right]^{-\frac{1}{2}} dz$. If a source has no available redshift, the median value of the corresponding group is adopted for the calculation. For the FSRQs, relevant radio information was obtained from Browne and Murphy (1987) and Fan et al. (2011). The final sample consists of 513 blazars with complete data at 5 GHz, which include 250 BL Lacs and 263 FSRQs. Spectral classifications of these objects is available in Pei et al. (2020a; 2020b). Among the BL Lacs are 114 HSPs, 50 ISPs and 86 LSPs. For a complete discussion of the unified scheme, we calculate and include the γ-ray and radio luminosities of 340 non *Fermi*-detected radio galaxies of Pei et al (2020b) and Pei et al (2019) respectively. All through the paper, we adopt the standard cold dark matter (Λ-CDM) cosmology with Hubble's constant $H_0$ = 72 kms$^{-1}$Mpc$^{-1}$ and $\Omega_0 = \Omega_m + \Omega_\Lambda$, ($\Omega_m$ = 0.30; $\Omega_\Lambda$ = 0.70). All relevant data are adjusted based on this concordance cosmology. For the statistical analyses, the degree of relationship between these emission parameters is inferred from the Pearson Product Moment correlation coefficient (*r*) using python and MatLab programming softwares

**3.1 Distributions of observed radio source parameters**

The distributive analyses of the observed source parameters of our sample are particularly necessary in investigating unification scheme so as to understand the relationship between the various classes and subclasses of these sources. For purposes of unification studies, these parameters are examined separately in radio galaxies, BL Lacs and FSRQs. The distribution of core-dominance parameter (*C*) and the cumulative probability of our sample on logarithmic scale is shown respectively in the upper and lower panels of Figure 1. Actually, the distribution is continuous, with the radio galaxies displaced to the lowest core-dominance parameter, ISPs and LSPs appear to have larger values of *C* in the data while FSRQs spread over the entire range; indicating that some ISPs and LSPs are more core-dominated than FSRQs. A cursory look at the figure indicates that FSRQs and radio galaxies fairly show log normal distributions in a unimodal configuration while HSPs, ISPs and LSPs show rather flat distributions. Nevertheless, the distributions yield mean (logarithm) values ~ 0.62 ± 0.20 for radio galaxies, 0.54 ± 0.10 for HSPs, 1.44 ± 0.30 for ISPs, 1.42 ± 0.20 for LSPs and 0.68 ± 0.10 for FSRQs. Using $R_T = 0.024$, which appears to be consistent with unification of blazars and radio galaxies (Orr and Browne, 1982; Pei et al. 2020b), the angle to the line of sight of each source sample was calculated from equation (4). The results yield 13.5° for LSPs, 14.8° for ISPs, 16.8° for FSRQs, 20.4° for HSPs and 28.2° for radio galaxies. These results are consistent with the predictions of the unified scheme in which FSRQs and now, LSPs and ISPs are strongly beamed and are inclined at smaller angles than radio galaxies and HSPs. Thus on average, it can be argued that LSPs and ISPs are systematically more core-dominated than FSRQs, which are in turn, more core-dominated than radio galaxies, with HSPs, on the other hand showing the lowest core-dominance. **However, relativistic beaming model predicts that more core-dominated sources should possess larger Doppler boosting factors (Pei et al., 2020d), which are found to be larger for LSPs and ISPs than for HSPs (e.g. Nieppola et al., 2008). This result seems to suggest that while LSPs and ISPs could represent extremely beamed counterparts of FSRQs, HSPs are intrinsically a different subclass of BL Lacs. Thus, in the frame work of the popular orientation-based unification, misaligned blazars will drop in luminosity and Doppler boosting with increasing viewing angle (e.g. Meyer et al. 2011). The lowest core-dominance of HSPs can be interpreted to mean that HSPs are more misaligned versions of BL Lac sources.** Nevertheless, a Kolmogorov-Smirnov (*K-S*) test between the distributions of log *C* for radio galaxies, FSRQs and individual subclasses of BL Lacs shows that although the observed average sequence subsists, there is a considerable overlap, the chance probability $p < 10^{-4}$ at 95%. This suggests that there is no significant difference between the underlying distributions of the objects in log $C$.

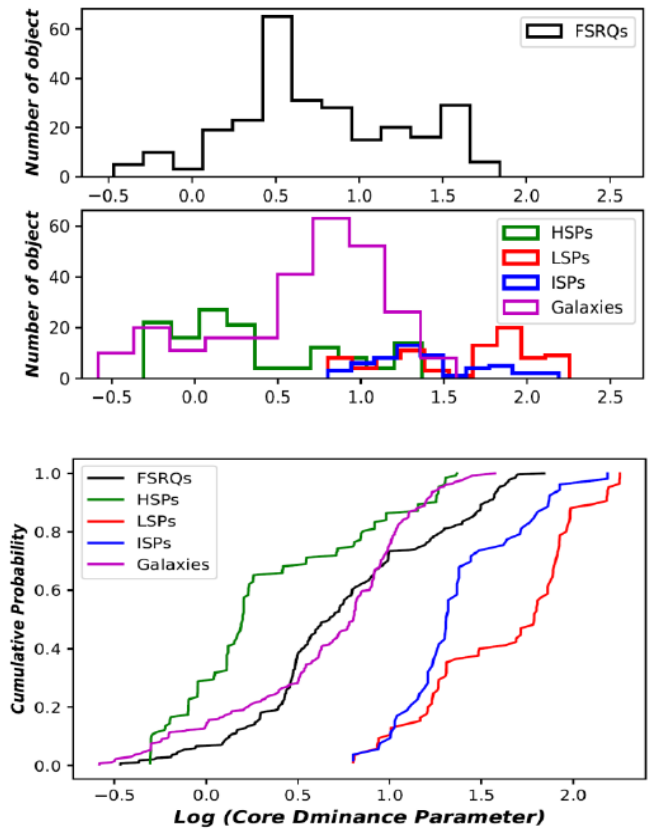

**Figure 1: Distribution of core-dominance parameter (upper panel) and cumulative probability (lower panel) of our sample**

For the calculated luminosities, we use the distributions to investigate the differences in extended and core luminosities between the different subclasses of our sample. Though, the extended luminosity of radio sources is understood to be isotropic, orientation invariant, and not very significant in explaining the beaming properties of extragalactic radio sources (see, e.g. Urry and Padovani, 1995; Kollgaard et al., 1996), for the purpose of comparison with other observed properties of the core-dominated sources, we show the histogram distributions on logarithmic scales for the present sample in Fig. 2a (upper panel). It is clear from the figure that FSRQs, on average, have the tendency to possess higher values of extended luminosity than the BL Lacs and radio galaxies. The average values are 26.60 ± 0.20, 25.50 ± 0.10, 25.12 ± 0.10, 24.30 ± 0.20 and 22.80 ± 0.20 for FSRQs, LSPs, ISPs, HSPs and radio galaxies, respectively. Hence, the average values of the extended luminosities (**W·Hz⁻¹**) for the subsamples following the relation: $\langle L_E \rangle|_{\text{GALAXY}} < \langle L_E \rangle|_{\text{HSPs}} < \langle L_E \rangle|_{\text{ISPs}} < \langle L_E \rangle|_{\text{LSPs}} < \langle L_E \rangle_{\text{FSRQs}}$. The distribution is consistent with the galaxy - BL Lac - FSRQ- unification with LSPs and ISPs showing overlapping distributions. However, the multimodal distribution of these objects arises from the fact that there are different luminosity values for these sources. This result can be interpreted to mean that in terms of extended luminosity, different subclasses of BL Lacs may exist and that extended luminosity can be used to investigate the blazar sequence (Pei et al., 2019).

In the case of core luminosity, radio galaxies, BL Lacs and FSRQs apparently do not show any log-normal distribution but rather have bimodal configuration for radio galaxies and HSPs as shown in Fig. 2b. FSRQs range from 26.08 to 29.80 with peak at 28.50 and an average value of 28.18 ± 0.20. However, BL Lacs range from 23.10 to 28.40 peaking at different values for HSP, LSP and ISP subclasses while radio galaxies range from 22.50 to 25.90 with peaks at 23.20 and 24.20. The average values of BL Lac subclasses are 26.40 ± 0.20, 26.84 ± 0.10, and 27.11 ± 0.10 for LSPs, ISPs and HSPs, respectively and 24.39 ± 0.20 for radio galaxies. We find that the average values of the core luminosity (**W·Hz⁻¹**) of our sample following the relation: $\langle L_C \rangle|_{\text{GALAXY}} > \langle L_C \rangle|_{\text{HSPs}} > \langle L_C \rangle|_{\text{ISPs}} > \langle L_C \rangle|_{\text{LSPs}} > \langle L_C \rangle|_{\text{FSRQs}}$, a trend that is more or less a mirror image of the extended luminosity distribution, which suggests a sequence for our sample. It is observed from the distribution that LSPs, ISPs and FSRQs are displaced to higher values of the core luminosity. This is as expected since the simplest relativistic hypothesis predicts that sources inclined at small angles to the line of sight should have more prominent cores. Hence, the observed displacement in the radio core power for these radio sources is consistent with unified scheme. However, the distributions of radio galaxies, LSPs, ISPs and FSRQs in $L_E$ and $L_C$ are observed to be continuous with no distinct dichotomy between the subclasses in such a way that is in agreement with the unification scheme for these sources, signifying that the populations of the sample are inherently related. A two sample *K–S* test was carried out on both extended and core luminosities. The cumulative distribution probabilities of the different subclasses are shown in Figures 2 (a,b lower panels) respectively. It is found that for the core luminosity, the distribution shows that they belong to different parent populations ($p = 2.54 \times 10^{-27}$). For the extended luminosity, the overlap is very obvious for LSPs, ISPs and FSRQs, signifying that radio galaxies, HSPs and combined ISPs, LSPs and FSRQs are statistically different in extended luminosity. However, there is a sequence in the distribution of both extended and core luminosity in a sense that is consistent with a blazar sequence that can be extended to the radio galaxies.

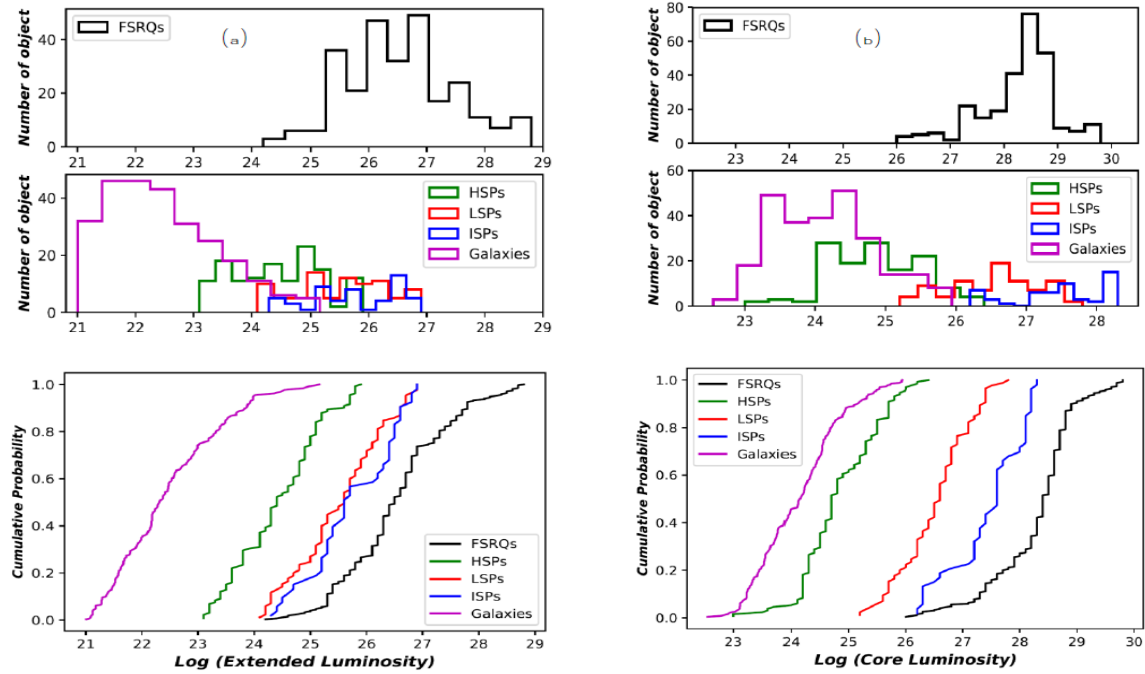

**Figure 2: Distribution of (a) extended luminosity (upper panel) and cumulative probability (lower panel) (b) core luminosity (upper panel) and cumulative probability (lower panel) of our sample**

We also show the distributions of the objects in γ-ray luminosity **(erg s$^{-1}$)** in Fig. 3. Apparently, the distribution is continuous, with the FSRQs displaced to highest *γ*-ray luminosity and radio galaxies to lowest γ-ray luminosity, suggesting that FSRQs are stronger γ-ray emitters than BL Lacs and radio galaxies. Nonetheless, the distributions yield mean (logarithm) values ~ 46.70 ± 0.30 for FSRQ, 45.80 ± 0.20 for LSP, 45.30 ± 0.20 for ISP, 44.30 ± 0.20 for HSP and 43.20 ± 0.30 for galaxies, which follow the sequence $Log\langle L_\gamma \rangle |_{FSRQs} > Log\langle L_\gamma \rangle |_{LSPs} > Log\langle L_\gamma \rangle |_{ISPs} > Log\langle L_\gamma \rangle |_{HSPs} > Log\langle L_\gamma \rangle |_{GALAXIES}$ suggesting that the blazar sequence can extend to jetted radio galaxies. Statistical test reveals that FSRQs and BL Lac subclasses are nicely fitted to a log-normal distribution with skewness (*μ*) in the range - 0.03 ≤ μ ≤ 0.05. The observation suggests that similar mechanisms are responsible for the variations in γ-ray luminosity of all subclasses of blazars. Furthermore, two-sample *K-S* test carried out on the data shows that in general, at 95% confidence there is almost a zero probability ($\rho < 10^{-8}$) that there is any fundamental difference between the underlying distributions of these objects in γ-ray luminosity. Similarly, the radio galaxies do not appear to be significantly different from FSRQs and BL Lac subclasses and occupy the lowest luminosity regime of the distribution, which suggests that in general, there is a connection between FSRQs, BL Lacs and radio galaxies.

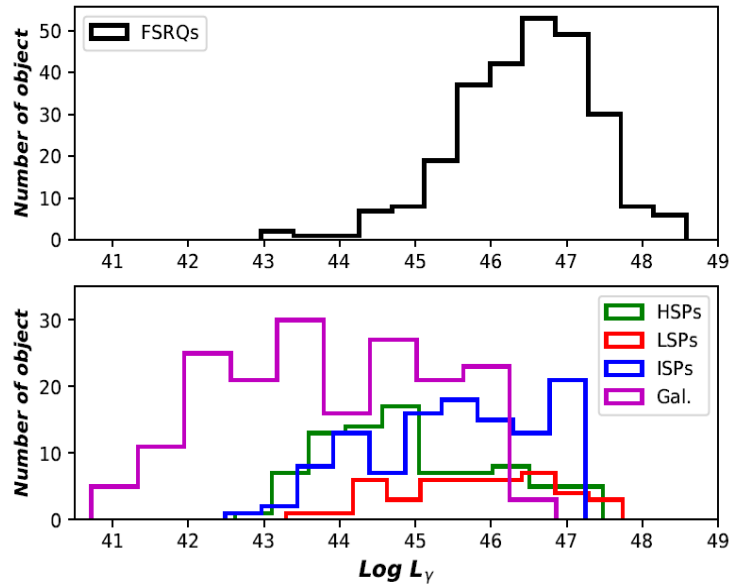

**Fig 3: Distributions of γ-ray luminosity of FSRQs, BL Lacs and radio galaxies**

### 3.2: The *C - L* relations and the blazar sequence

To investigate the effect of relativistic beaming on the evolution of γ-ray emission of blazars, we show in Fig. 4, the scatter plot of *C* against $L_\gamma$ for the current sample. There is a tendency for the sample to show a positive trend on the *C* − $L_\gamma$ plane with *r* ~ 0.30. The positive trend can be interpreted to mean that γ-ray emission in AGNs is also beamed. Nevertheless, this positive trend only exists when the entire AGNs are taken together. For each individual subclass of the sample, there is no discernible trend. This implies that relativistic beaming alone cannot explain the variations in γ-ray emission of blazars.

13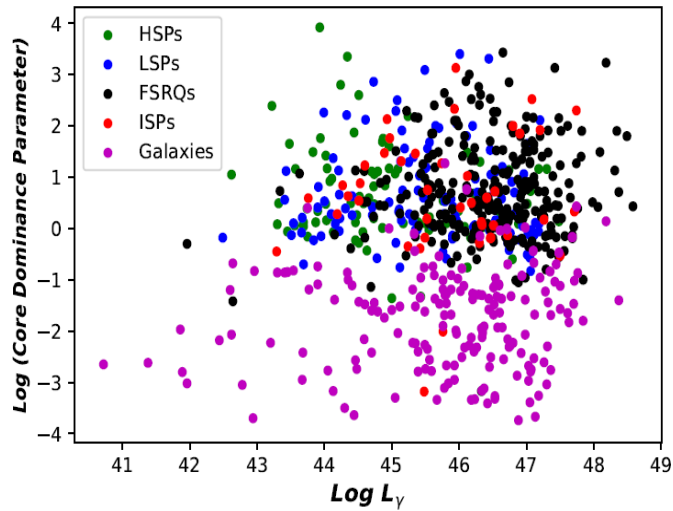

**Figure 4: Plot of radio core-dominance parameter against γ-ray luminosity of blazars and radio galaxies**

To examine the relative effects of relativistic beaming in the sample, the scatter plots of the core-dominance parameter as a function of extended and core luminosities on logarithmic scales are shown in Figure 5. For the $C - L_E$ data in Fig. 5 (a), FSRQs, ISP- and LSP- BL Lacs overlap in a sense that is consistent with a unified scheme via relativistic beaming model. However, three distinct groups of objects can be identified, namely, radio galaxies, HSP- BL Lac and the combined class of ISP, LSP- BL Lacs and FSRQs. There is a clear anti-correlation of $C - L_E$ data for each of the groups. Table 1 shows the results of regression analysis of the subsamples. The slope $k$, intersection $k_0$, correlation coefficient $r$ and chance probability $p$ and their errors are all listed in the table.

**Table 1: Results of linear regression fitting given as:** $y = (k \pm \Delta k)x + (k_0 \pm \Delta k_0)$

| plots | *Sample* | *K* | Δk | $k_0$ | Δ$k_0$ | *r* | *p* |
|---|---|---|---|---|---|---|---|
| $\log C - \log L_E$ | radio galaxies | -0.93 | 0.30 | 1.27 | 0.10 | -0.65 | $< 10^{-4}$ |
| $\log C - \log L_E$ | HSP-BL Lac | -0.42 | 0.10 | 1.09 | 0.20 | -0.62 | $< 10^{-4}$ |
| $\log C - \log L_E$ | combined ISPs, LSPs and FSRQs | -0.56 | 0.20 | 2.90 | 0.10 | -0.67 | $< 10^{-4}$ |

These groups are somewhat parallel on the $C - L_E$ plane with a form of scaling factor on $L_E$. We interpret this to mean that similar processes give rise to the $C - L_E$ anti-correlation in the different groups at intrinsically different scales. The results suggest that while some BL Lacs (ISPs and LSPs) can be extremely beamed FSRQs, there is a fundamental difference between HSPs and other subclasses of BL Lacs. Hence, it can be reasoned that the observed dichotomy between the radio galaxies, HSP-BL Lac and combined ISPs, LSPs and FSRQs in extended luminosity is as a result of the differences in some intrinsic properties that vary in a sequence across the samples, which is consistent with a blazar sequence.

However, while the core-dominance parameter is strongly anti-correlated with extended luminosity for radio galaxies, HSP-BL Lac and combined ISPs, LSPs and FSRQs, as shown in Fig. 5(a), that of core luminosity shown in Fig. 5(b) is less important. Table 2 shows the results of regression analysis of the subsamples.

**Table 2: Results of linear regression fitting given as:** $y = (k \pm \Delta k)x + (k_0 \pm \Delta k_0)$

| plots | *Sample* | k | Δk | k₀ | Δk₀ | r | p |
|---|---|---|---|---|---|---|---|
| $\log C - \log L_C$ | radio galaxies | -0.31 | 0.20 | 0.37 | 0.20 | -0.35 | $< 10^{-4}$ |
| $\log C - \log L_C$ | HSP-BL Lac | -0.52 | 0.10 | 1.07 | 0.20 | -0.21 | $< 10^{-4}$ |
| $\log C - \log L_C$ | combined ISPs, LSPs and FSRQs | -0.34 | 0.20 | 1.03 | 0.10 | -0.47 | $< 10^{-4}$ |

It can be argued that the strong $C - L_E$ anti-correlation observed in this study arises due to redshift effects on the parameters (cf. eq. 9). To clear out such doubts, we eliminate the effect of redshift on these parameters by subtracting out the common dependence of both $C$ and $L_E$ on redshift using Spearmann's partial correlation analyses (e.g. Padovani et al. 1992; Odo and Ubachukwu, 2013) given as

$$r_{12,3} = \frac{r_{12} - r_{13}r_{23}}{\sqrt{(1-r_{13}^2)(1-r_{23}^2)}} \tag{10}$$

where $r_{12}$ represents the correlation coefficient between $x_i$ and $x_j$, and $r_{ij,k}$ is the partial correlation coefficient between $xi$ and $xj$, with $z$ dependence removed; ($i, j, k = 1,2,3$). Simple regression analyses of current data yield $r$ - 0.65 and - 0.56 for $L_E - z$ and $C - z$ data respectively, for the current sample. Applying this to current $C - L_E$ data independent of $z$ yields, $r_{12,z}$ = - 0.64, - 0.61, - 0.68 for radio galaxies, HSP-BL Lac and combined sample of ISPs, LSPs and FSRQs respectively. Similar analyses for $C - L_C$ data did not yield any significantly different results: $r_{12,z}$ = - 0.33 for radio galaxies, - 0.23 for HSP-BL Lac and - 0.46 for combined sample of ISPs, LSPs and FSRQs ($p \sim 0$ in all the cases). Thus, we argue that the correlations observed are independent of cosmological effects. These correlations may therefore be intrinsic, rather than an artifact of redshift in the present sample of these radio sources.

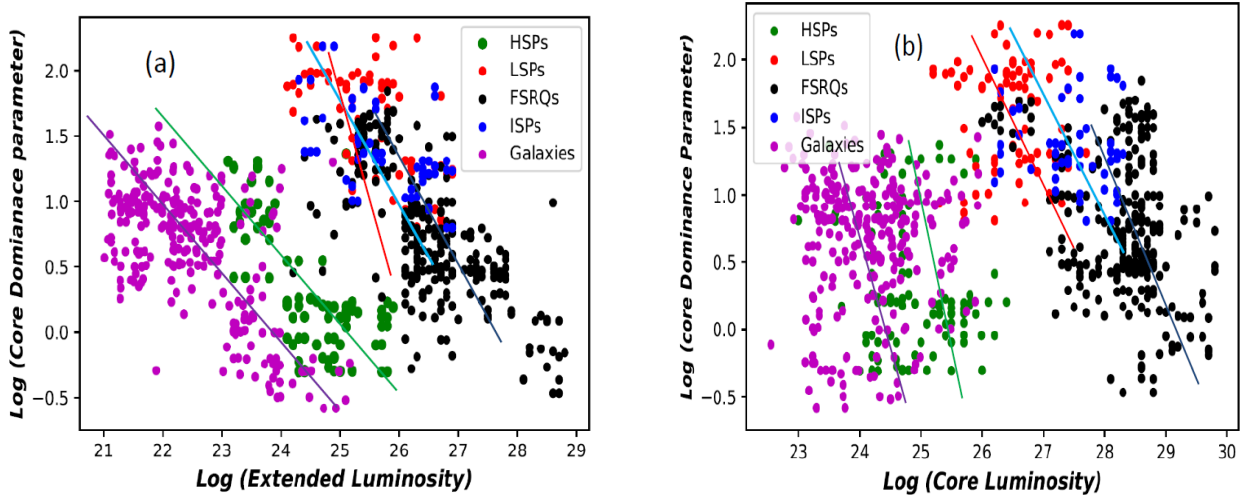

**Figure 5: Scatter plots of core-dominance parameter as a function of (a) extended luminosity (b) core luminosity on logarithmic scales for our sample**

### 3.3: Correlations between $C$ and Redshift

A redshift-dependent core dominance parameter of radio sources of the form $C = \left(\frac{L_C}{L_E}\right)(1+z)^{\alpha_C - \alpha_E}$ has been widely predicted to explain the unification of extragalactic radio sources (e.g. Fan and Zhang, 2003). We examine the implications of this redshift effect on the present sample in a considerable detail. Figure 6 shows the scatter plot of $C - z$ on logarithmic scales for our sample. The spread in redshift is significantly smaller for radio galaxies and HSP-BL Lac than for combined sample of ISPs, LSPs and FSRQs. However, the connection between ISP, LSP and FSRQs is still prominent on the $C - z$ plane, with HSP-BL Lac being significantly different from ISPs and LSPs. This again suggests that ISPs, LSPs and FSRQs can be unified through luminosity evolution and that there are intrinsic differences between HSPs and other subclasses of BL Lacs. There is a significant $C - z$ anti-correlation ($r \sim -0.58$) for the combined sample of ISPs, LSPs and FSRQs. Nevertheless, little or no correlation ($r \sim -0.39$) exists for HSP-BL Lac and radio galaxies. It is important to observe that all the radio galaxies and HSP-BL Lac are located at low redshifts. Perhaps, for these low redshift sources, evolutionary effects may be less important.

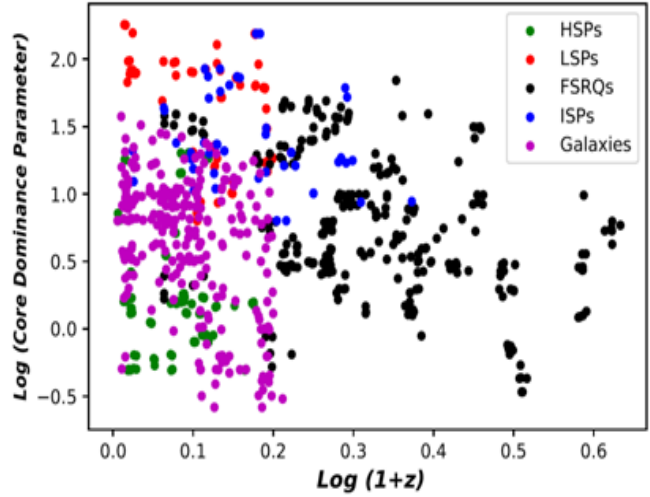

**Figure 5: Log –Log scatter plot of radio core-dominance parameter against redshift of our sample**

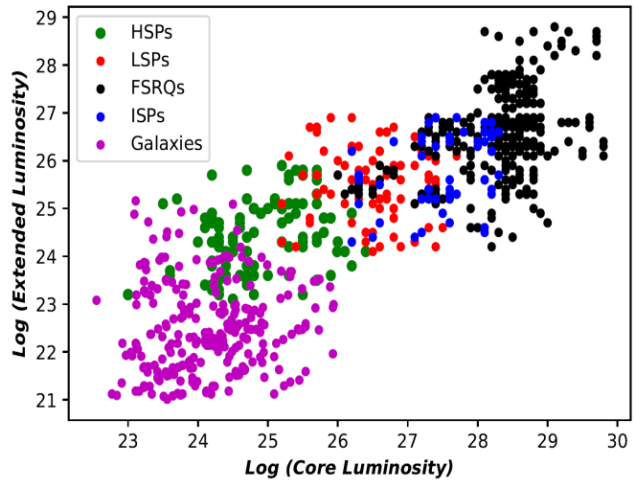

**Figure 6: Log –Log scatter plot of radio extended luminosity against core luminosity of our sample**

## 4. Discussion

We have used relativistic beaming model to investigate the relationship between FSRQs, BL Lacs and radio galaxies in a sample of F*ermi*-LAT AGNs. It is widely known that blazars form an exceptional population of AGNs with similar emission processes effective under a variety of physical conditions. Blazar sequence unification scheme proposes that both FSRQs and BL Lac objects are forms of objects with the same physical process but differ in bolometric luminosity (Fossati et al. 1997; 1998; Ghisellini et al. 1998; Abdo et al. 2010). In addition, FSRQs, BL Lacs and radio galaxies represent progressively misaligned populations of AGNs. Thus, there must be no division in the luminosity distributions of these sources.

Remarkably, our results show from the comparison of the core and extended luminosity that the distributions of the present sample evidently indicate that ISPs and LSPs could be extreme version of FSRQ populations. This appears to be in agreement with an earlier supposition that FSRQs can evolve into BL Lacs, becoming weak-lined objects by virtue of increased beaming of the continuum (Vagnetti et al., 1991) suggesting that BL Lacs are FSRQs with highly boosted continuum. We investigated the angle of sight for each subsample and found that ISPs and LSPs subclasses are inclined at sharper angles to the line of sight than HSPs and radio galaxies. Hence, at small angles to the line of sight, relativistic beaming is more prominent in LSP and ISP-BL Lacs. Meanwhile, in the simplest unification scheme for radio sources, ISP, LSP and FSRQs are presumed to be the extreme beamed counterparts of radio galaxies that form the unbeamed parent populations. So, the difference between the core and extended luminosity is supposed to be apparent, which led to high values of the core-dominance parameter for the ISPs and LSPs.

Another important outcome of current investigation is the strong anti-correlation between the core-dominance parameter and extended luminosity for the different subclasses of our sample. It has been shown that HSP, radio galaxies and FSRQs with ISPs and LSPs occupy parallel regions on the core-dominance – extended luminosity ($C$ – $L_E$) plane. It can thus be debated that similar mechanisms are responsible for their variation with some form of a scaling factor on $L_E$. Thus, there is a sequence IN this variation from radio galaxies to FSRQs in a sense that seems somewhat consistent with blazar sequence, which posits that different blazar subclasses are different manifestations of the same physical process that differ only by bolometric luminosity (Fossati et al. 1998; Ghisellini et al. 1998). However, results from recent investigations (e.g.

Rani, 2019; Iyida et al. 2020) suggest that the blazar sequence is a continuum that can be extended to the jetted radio galaxies, which form the parent populations of blazars. The region of the $C - L_E$ plot in Fig. 5 occupied by the radio galaxies, relative to blazar subclasses, is in good agreement with the blazar sequence extending to radio galaxies and suggests that similar mechanisms are responsible for the $C - L_E$ variations, with a scaling factor. Furthermore, the $C - L_E$ trend was observed to be negative for the individual subsamples of radio galaxies, HSPs and the combined sample of ISP, LSP and FSRQs, which is in agreement with the beaming model (Orr and Browne, 1982; Odo et al., 2012). It is therefore expected that relativistic beaming effects would be well prominent in the combined sample of ISP, LSP and FSRQs. Thus, these results are quite consistent with a number of previous results on beaming effects in AGN sources (e.g., Zhou et al., 2007; Fan et al. 2010; 2011; Pei et al. 2019).

The fact that HSPs, radio galaxies and combined sample of ISP, LSP and FSRQs occupy discrete and nearly like sections in the scatter plot of core-dominance parameter – luminosity implies that these AGN types may have distinctive evolution and histories (e.g. Urry and Padovani, 1995; Pei et al. 2019), thus, can be unified via blazar sequence. On the other hand, these observations seem to suggest that HSPs are fundamentally different from other subclasses of BL Lacs rather than beaming or evolutionary connections. In fact, Odo et al., (2017) argued for a dichotomy between radio and X-ray selected (RBLs and XBLs respectively) BL Lacs using the 1 Jy, HEAO LASS and EMSS samples of BL Lacertae objects. Perhaps, the difference between HSPs and other BL Lac samples reported in current investigation is a manifestation of the XBL/RBL dichotomy. Obviously, most of the XBL objects in Odo et al., (2017) are HSP sources, while all the RBLs are ISPs and LSPs. The differences between the BL Lac subclasses can be attributed to intrinsically different rates of cooling at different positions within the cores of the various subclasses of blazars (e.g. Ghisellini et al. 1998; Savolainen et al. 2010). However, the distribution of the objects in each subclass of AGNs on the $C - L_E$ plane is attributable to beaming effects such that these objects may vary in orientation (Li et al. 2016; Resconi et al. 2017; Alhassan et al. 2019; Iyida et al. 2020).

Perhaps, it can be maintained that the $C - L_E$ anti-correlation observed in this analysis could be an observational effect connected with the difficulty in detecting weak extended flux around strong point sources. To investigate this observational effect, we plot the scatter of $L_E$ as a function of $L_C$ for the entire sample in Fig. 7. Apparently, there is a strong indication that $L_E$ scales positively with $L_C$ ($r \sim 0.76$), which is not in agreement with the observational effect associated with dynamical range of the sources. In fact, analyses of radio data of traditional BL Lacs (Odo and Ubachukwu, 2013) and core-dominated quasars (Odo et al., 2015) suggest that radio sources projected close to the line of sight can have $L_E$ beamed relative to $L_C$. If this is actually the case, the observed anti-correlation can be understood in the framework of relativistic beaming if beaming persists at largest scales. It is also arguable that the $C - L_E$ anti-correlation in the sample may have arisen as a result of redshift effect on the parameters. Nevertheless, the correlation persists after subtracting out the common dependence of $C$ and $L_E$ on redshift using Spearman's partial correlation statistic ($\rho$), which can be interpreted to mean that the $C - L_E$ anti-correlation is not a secondary effect of redshift dependence of the parameters.

It is evident from current investigation that the blazar subclasses can be explained very well in the framework of relativistic beaming effect such that ISP, LSP and FSRQs are the beamed counterparts of both Fanaroff-Riley I (FR I) and Fanaroff-Riley II (FR II) respectively (Ghisellini et al. 2009; Sbarrato et al. 2014). Nevertheless, the presence of high radio power BL Lacs and low power FSRQs have been noted in various earlier blazar samples (Kharb et al., 2010; Landt et al., 2006; Caccianiga and Marchã 2004; Padovani et al., 2003; Perlman et al., 1998;), which perhaps, breaks the simple dichotomy between the two Fanaroff-Riley categories of AGN. Moreso, Giommi *et al.* (2013) proposed that a large percentage (over 70%) of ISPs and LSPs from the *Fermi*-LAT have strong, but hidden, emission lines and thus, can be classified as FSRQs. The prediction has been substantially supported (Ackermann et al. 2015; Padovani et al., 2019; Odo and Aroh, 2020) with the detection of a number of intrinsically *Fermi*-LAT FSRQs masquerading as BL Lacs, with γ-ray properties departing significantly from that expected from traditional FSRQs by several orders of magnitude. Furthermore, it has been observed from analysis of third *Fermi*-LAT catalogue (Ackermann et al., 2015) that quite a number of *Fermi*-LAT sources were in flaring state most of the time, which could make their bolometric flux much higher than average at such times. Perhaps, the observed flux of ISP and LSP-BL Lacs in current analysis could have arisen if they were in flaring state during the duty cycle which data is presented in the **fourth catalogue of *Fermi*-LAT (Abdollahi et al., 2020; Ajello et al., 2020)**. However, the order in which these ISP- and LSP-BL Lacs appear in the current study is consistent with an orientation-based unification scheme, suggesting that there might be a kind of unification scheme for these objects and FSRQs. Nevertheless, for the BL Lacs, the clear separation of HSPs and LSPs on the $C - L_E$ plane which is not filled by the ISPs can be interpreted in terms of a fundamental difference between them.

## 5. Conclusion.

We have investigated the distributions and correlations between core-dominance parameter and both extended and core luminosities of some *Fermi* selected blazars to quantitatively test for relativistic beaming in the light of the blazar sequence. Our results revealed that there is a sequence from radio galaxies through BL Lac subclasses to FSRQs, with ISPs, LSPs and FSRQs being strongly beamed than their counterparts, which is in agreement with radio galaxies –BL Lac -FSRQs unification and strongly supports the earlier supposition of a unified scheme for blazars and radio galaxies. We find a fundamental difference between HSP BL-Lacs and other subclasses of BL Lacs. There is a strong anti-correlation between core-dominance parameter and extended luminosity in each subclass of the AGNs which suggests that relativistic beaming persists at largest scales. We observed weak correlation between $C$ and redshift for the individual subsamples which is indicative that $C$ does not strictly depend on redshift, but, perhaps, on other possible factors.

## Acknowledgement

We sincerely thank an anonymous referee whose invaluable comments and suggestions helped to improve the manuscript. This work is financially supported by the National Economic Empowerment and Development Strategy (NEEDS) of the Federal Government of Nigeria and partly done at the North-West University, Potchefstroom, South Africa.

**REFERENCE**
Abdo, A. A., Ackermann, M., Ajello, M., Atwood, W. B. *et al.* 2009, *ApJ,* 700, 597
Abdo, A. A., Ackermann, M., Ajello, M., Atwood, W.B. et al. 2010, *ApJ*, 710,1271
Abdollahi, S., Acero, F., Ackermann, M., et al. 2020, ApJS, 247, 33,
Ackermann, M., Ajello, M., Allafort, A., *et al.* 2011, *Astrophysical Journa*, 743, 171
Ackermann M., Ajello, M., Atwood, W.B., Baldri, E. *et al.* 2015, *Astrophys. J.* 810: 14-47
Ajello, M., Angioni, R., Axelsson, M., et al. 2020, ApJ, 892, 105,

Alhassan, J.A., Ubachukwu, A.A. & Odo, F.C. 2013, *JoAA,* 34, 61

Alhassan, J.A., Ubachukwu, A.A., Odo, F.C and Onuchukwu. C.C. 2019, *RMxAA,* 55, 151-159

Andruchow, I., Romero, G.E., & Cellone, S.A., 2005, Astro. & Astro., 442, 97

Angelakis, E., Fuhrmann, L., Marchili, N., Foschini, L. *et al.* 2015, *A&A.* 575: A55

Antonnuci, R. 1993, *A &A* 31, 471

Balokovic, M., Smolcic, V., Ivezic, Z. & Zamorani, G. 2012. *ApJ.* 759:30

Bicknell, G. V. 1995, *ApJS,*101, 29

Bridle, A.H., Stanghellini, C., Lonsdale, C.J. Burns, J and Laing, R.A. 1994, *ApJ*, 108. 766

Browne I. W. and Murphy, D. W., 1987, *MNRAS*, 226, 601

Caccianiga, A. and Marchã, M. J. 2004*, MNRAS*, 348, 937

Chen, Y. Y., Zhang, X., Xiong, D. R., Wang, S. J. and Yu, X. L. 2016, *Res. Astro. Astrophy.*16, 13

D'Abrusco, R., Crespo, N. A., Massaro, F. *et al*. 2019, *ApJS*, 242,1

Deenert-Thrope, J, Barthel, P.D. and Van Bmmel, I.M. 2000, *A & A*, 364, 501

Fan, J. H., Liu, Y., Li, Y., et al. 2011, *JoAA*, 32, 67

Fan, J. H., Yang, J., Tao, J., Huang, Y. and Liu, Y. 2010, *PASJ*, 62, 211

Fan, J.H. and Zhang, J.S. 2003, *JoAA*, 407 (3), 899

Fan, J. H., Yang, J. H., Liu, Y., et al. 2016, ApJS, 226, 20

Fan, J., Yang, J., Yuan, Y., Wang, J. Gao, Y. 2012, *ApJ*, 761:125

Finke, J. D. 2013, *ApJ*, 763, 134

Fossati, G., Maraschi, L., Celotti, A., Comastri, A. and Ghisellini, G.1998, *MNRAS,* 299, 433

Ghisellini G., Tavecchio F., 2008, MNRAS, 387, 1669

Ghisellini G. & Celotti A., 2001, A&A, 371, L1

Ghisellini, G., Righi, C., Costamante, L. and Tavecchio, F. 2017, *MNRAS,* 469, 255

Ghisellini, G., Celotti, A., Fossati, G., Maraschi, L., Comastri, A. 1998, *MNRAS,* 301, 451

Ghisellini, G., Maraschi, L. & Tavecchio, F. 2009, MNRAS, 396, L105

Ghisellini, G., Tavecchio, F., Foschini, L., et al. 2010, MNRAS, 402, 497

Ghisellini G., Tavecchio F., Foschini L, Ghirlanda G., 2011, MNRAS, 414, 2674

Giommi, P., Padovani, P. and Polenta, G. 2013. *MNRAS,* 431, 1914

**Hovatta, T., Valtaoja, E., Tornikoski, M. and Lahteenmaki, A. 2009, A &A, 494, 527**

Iyida, E.U., Odo F.C. Chukwude, A.E. and Ubachukwu, A.A 2019, *PASN*; 4, 74–81

Iyida, E.U. Odo, F.C., Chukwude, A.E. and Ubachukwu, A.A. 2020: *Open. Astro*. 29: 168-178

Kharb, P., Lister, M. L. and Cooper, N. J. 2010, *ApJ*. 710, 764

Kollgaard R. I., Palma C., Laurent-Mueheisen S. A. and Feigelson E. D. 1996, *ApJ*, 465, 115

Jeyakumar, S., Wiita, P.J., Saikia, D.J. and Hooda, J.S. 2005, *A&A*,432(3), 823

Li, Y., Fu, S. Y., Feng, H. J., He, S. L. et al., 2016, *Astrophys. & Astron.*38, 22

Liu Z., Wu Z. and Gu M. 2016, Res. Astron. Astrophys., 16, 103

Lind K. R. and Blandford R. D.1985. *ApJ*. 295, 358

Lobanov, A.P., Krichbaum, T.P., Witzel, A and Zeus, J.A. 2006, *PASJ*. 58(2). 253

Mao, P., Urry, C. M., Massaro, F., Paggi, A., Joe, C. et al., 2016, *ApJS*., 224, 26

Meyer E. T., Fossati G., Georganopoulos M. and Lister M. L. 2011, *ApJ*,740, 98

Nalewajko, K. and Gupta, M. 2017. *Astro &Astro*, 606, A44

**Nieppola, E., Valtaoja, E., Tornikoski, M., Hovatta, T. and Kotiranta, M. 2008, *A&A*. 488, 867**

Odo F. C., Chukwude, A. E. and Ubachukwu, A. A. 2017, *Astrophys. Space Sci.* 362, 23

Odo F.C., Chukwude A.E. and Ubachukwu, A.A. 2014, *Astrophys. Space Sci.* 349, 939

Odo F. C., Ubachukwu A. A., Chukwude A. E. 2015, *Astrophys. Space Sci.* 357, 147

Odo F.C. and Ubachukwu A.A. 2013, *Astrophys. Space Sci.* 347(2), 357

Odo F. C., Ubachukwu A. A., Chukwude A. E. 2012. *Journal of Astrophy. Astron. 33, 279*

Orr, M. J. and Browne, I. W. A. *MNRAS,* 1982,200, 1067

Padovani, P. 2007, *Astrophys. Space Sci.,* 309, 63

Padovani, P., Perlman, E. S., Landt, H., Giommi, P. and Perri, M. 2003, *ApJ,*588, 128

Padovani, P. 1992, *MNRAS,* 257, 404

Palladino, A., Rodrigues, X., Gao, S. and Winter, W. 2019, *ApJ,* 871; 41

Pei Z., Fan J., Yang J. and Bastieri D. 2020a, Publ. of the Astron. Soc. of the Pacific, 132:114102

Pei Z., Fan J., Bastieri D., Yang J., Xiao H., 2020b, *Science China. Phy., Mech., and Astr.*, 63, 259511

Pei, Z., Fan, J. H., Bastieri, D. and Yang J H. 2020c. *Res. Astron. Astrophy*. 20, 25

Pei, Z.,Fan, J. H., Bastieri, D., Sawangwit, U. and Yang, J H. 2019. *Res. Astron. Astrophy*. 19, 70

**Pei, Z., Fan, J.H. Yang, J.H., and Bastieri, D. 2020d, *PASA*, 37, 43**

Perlman, E., Padovani, P., Giommi, P., et al. 1998, *Astronomical Journal*, 115, 1253

Qin, P. and Xie, G.Z. 1998. *Astron. Astrophys. Suppl. Series.* 133, 217

Raiteri, C. M. and Capetti, A. 2016, *A&A*, 587, A8

Resconi, E., Coenders, S., Padovani, P., Giommi, P. and Caccianga, L. 2017, MNRAS, 468,597

Savolainen, T., Homan, D.C., Hovatta, T. et al. 2010. *Jour. of Astro & Astro*. 512, A24

Sbarrato, T., Padovani, P., &Ghisellini, G. 2014, MNRAS, 445, 81

Scheuer, P.A.G. and Readhead, A.C. 1979. *Nature*. 277, 182

Ubachukwu, A.A. 2002, *Astrophys. Space Sci.* 279, 251

Urry, M. 2011, JApA, 32, 139

Urry, C.M., and Padovani, P. 1995, *PASP*, 107, 803

Vagnetti, F., Giallongo, E. and Cavaliere, A. 1991, *ApJ*. 368, 366

Wu, D., Fan, J., Liu, Y. et al., 2014. *Publ. Astron. Soc. Japan.* 66(6), 117

Xu, C., Livio, M. and Baum, S.,1999, Astronomical Journal. 118(3), 1169

Zhou, J., Fan, J., Li, J and Liu Y. 2007, *Chinese J. of Astron. &Astrophy*. 5, 629-638